%% file: main.tex
\documentclass[runningheads]{llncs}

\usepackage{hyperref}
\usepackage{booktabs}
\usepackage[T1]{fontenc}
\usepackage{graphicx}
\usepackage[capitalise,noabbrev]{cleveref}
\usepackage{multicol}
\usepackage{adjustbox}

\usepackage[
    disable,     %
    textwidth=.9\marginparwidth,
    colorinlistoftodos,
    prependcaption,
    linecolor=green,backgroundcolor=green!25,bordercolor=green
]{todonotes}
\usepackage{multicol}

\usepackage{multirow}
\usepackage{makecell}
\usepackage{tikz}

\DeclareRobustCommand\checker{%
    \scalebox{.5}{
        \begin{tikzpicture}%
            \node[draw, circle, fill=black] () {};
        \end{tikzpicture}%
    }
}

\DeclareRobustCommand\nocheck{%
  \scalebox{.5}{
    \begin{tikzpicture}
        \node[draw, circle] () {};
    \end{tikzpicture}%
  }
}

\DeclareRobustCommand\semicheck{%
  \scalebox{.5}{
    \begin{tikzpicture}
        \node[draw, circle] () {};
        \draw[draw,circle,fill=black] (0,4pt) -- (0,4pt) arc(90:270:.46em) --cycle;
    \end{tikzpicture}%
  }
}

\makeatletter
\if@todonotes@disabled
\else %
\paperwidth=\dimexpr \paperwidth + 6cm\relax
\oddsidemargin=\dimexpr\oddsidemargin + 3cm\relax
\evensidemargin=\dimexpr\evensidemargin + 3cm\relax
\marginparwidth=\dimexpr \marginparwidth + 3cm\relax
\let\oldtitle\title
\renewcommand{\title}[1]{\oldtitle{#1\(_{\colorbox{red!20}{\tiny disable todos}}\)}}
\fi
\makeatother

\begin{document}
\title{SAFARI: a Scalable Air-gapped Framework for Automated Ransomware Investigation}

\titlerunning{SAFARI}

\author{
    Tommaso Compagnucci\(^{1}\)    \and
    Franco Callegati\(^{1}\) \and
    Saverio Giallorenzo\(^{1,2}\) \and
    Andrea Melis\(^{1}\)        \and
    Simone Melloni\(^{3}\)      \and
    Alessandro Vannini\(^{1}\)
}

\authorrunning{Compagnucci et al.} %

\institute{\(^1\) Alma Mater Studiorum - Universit\`a di Bologna, Bologna, Italy\\
\(^2\)Olas Team, INRIA, Sophia Antipolis, France\\
\(^3\)ARPAE Emilia-Romagna, Italy}

\maketitle              %
\begin{abstract}
Ransomware poses a significant threat to individuals and organisations,
compelling tools to investigate its behaviour and the effectiveness of
mitigations. To answer this need, we present SAFARI, an open-source framework
designed for safe and efficient ransomware analysis. SAFARI's design emphasises
scalability, air-gapped security, and automation, democratising access to safe
ransomware investigation tools and fostering collaborative efforts. SAFARI
leverages virtualisation, Infrastructure-as-Code, and OS-agnostic task
automation to create isolated environments for controlled ransomware execution
and analysis. The framework enables researchers to profile ransomware behaviour
and evaluate mitigation strategies through automated, reproducible experiments.
We demonstrate SAFARI's capabilities by building a proof-of-concept
implementation and using it to run two case studies. The first analyses five
renowned ransomware strains (including WannaCry and LockBit) to identify their
encryption patterns and file targeting strategies. The second evaluates
Ranflood, a contrast tool which we use against three dangerous strains. Our
results provide insights into ransomware behaviour and the effectiveness of
countermeasures, showcasing SAFARI's potential to advance ransomware research
and defence development.

\keywords{Analysis Automation, Ransomware, Infrastructure as Code}
\end{abstract}

\input{intro-idea}

\input{technology}
\input{implementation}

\input{case_study}

\input{related_work}

\input{conclusion}

\section*{Acknowledgments}
We thank Matteo Cicognani for supporting the collaboration between ARPAE and
Università di Bologna. This study was carried out within the ``CYBER RANGE FOR
INDUSTRIAL SECURITY - CRI4.0 NELL'AMBITO DEL BANDO PER PROGETTI DI RICERCA
INDUSTRIALE STRATEGICA'' (PR FESR 2021--2027 AZIONE 1.1.2 CUP: E37G22000490007).

\bibliographystyle{splncs04}
\bibliography{biblio}

\end{document}

%% file: intro-idea.tex
\section{Introduction}
\label{sec:intro}
Ransomware is malware that seizes users' and organizations' data, demanding
payment to restore access to the rightful owners~\cite{LG16}.
Ransomware~\cite{RN17,OALU22} is one of the most pressing cybersecurity threats
facing organisations and individuals worldwide. In its most general definition,
ransomware extorts victims through their data. Economically, ransomware impose
sizable costs to society~\cite{HCCC20,MJTA22}. Indeed, ransomware forms a
sophisticated, multi-billion dollar industry that continues to adapt and
overcome defensive measures~\cite{MBS20}.

Despite increased awareness and investment in cybersecurity, ransomware attacks
continue to proliferate, causing significant financial losses, operational
disruptions, and potential harm to human life and safety~\cite{CB22}.

The study of the behaviour of ransomware in different contexts and conditions
provides fundamental knowledge for its contrast. The scientific and technical
literature abounds with solutions to counteract
ransomware~\cite{BMS18,MKCNW21,BBAHK21,OALU22,MSLXWLHNH24,dey2022daemon}.
Studying the behaviour of these solutions against real-world malware is
important to both evaluate their effectiveness and efficiency and help their
evolution.

We respond to such needs with a platform for the \textit{safe} and
\textit{efficient} investigation of ransomware and its mitigations. 
Safety and efficiency of investigation are the main challenges we address in the
design of our proposal. To perform an evaluation of ransomware attacks and of
contrast strategies, one needs numerous tests in which real ransomware attempts
to encrypt users' data. We provide a framework to run these tests
\textit{safely}, ensuring that the ransomware infection is fully contained, and
\textit{efficiently}, automating 
the management of the experiment parametrisation, to generate test batteries in
a consistent and reproducible manner, and to collect and analyse indicators
useful to understand the behaviour and estimate the effectiveness of ransomware
and defensive configurations.

\paragraph{SAFARI} We call our proposal Scalable,
Air-gapped Framework for Automated Ransomware Investigation (SAFARI), available
as an open-source project at \url{https://github.com/Flooding-against-Ransomware/SAFARI}.

One of the foundational principles behind SAFARI is the \emph{democratisation of
security research}. Indeed, while SAFARI's architecture is general and can
encompass the usage of Cloud virtual machines (VM), its design---and the
prototype we build to showcase it---targets on-premises deployments where
investigators can use local resources to assemble the system. In this way, small
research groups, as well as student collectives who can access
home-lab-sized hardware can configure their own deployment and run experiments.

Given the dangerous and highly infective nature of ransomware, SAFARI
run experiments in \emph{air-gapped} environments. Thanks to the
virtualisation of the testing environment, SAFARI isolates malware processes
within VMs from those of the hosting environment and the other VMs (to avoid
interference). 

SAFARI integrates complementary technologies for the definition of experiments
and the collection and analysis of attack data. 
Our proof-of-concept, presented in \cref{sec:implementation}, illustrates the
architecture of SAFARI both in terms of concepts and of software (off-the-shelf and purpose-made by us) that SAFARI integrates to analyse the behaviour of
ransomware and their contrast solutions.

While SAFARI's architecture can accommodate the study of malware in general
(combined with countermeasures), we focus our proposal on \textit{ransomware} of
the crypto kind, since it targets the data of the user, which we use as the
metric to measure the effectiveness of both malware and contrast tools. In
\cref{sec:CaseStudies}, we showcase SAFARI through two case studies. The first
investigates the behaviour and efficiency of real-world, infamous ransomware
such as WannaCry, LockBit, and Phobos. The second analyses the effectiveness and
efficiency of a recent tool, Ranflood~\cite{BGMMOP23,BGMMP24}, which contrasts
ransomware attacks through data flooding --- an innovative technique that mixes
dynamic honeypots and moving-target defence to contrast ransomware attacks.

Notably, while we provide the experimental data in \cref{sec:CaseStudies} as
evidence of the potential of SAFARI, the analysis itself constitutes a
contribution that adds new knowledge about the analysed malware's behaviour.
We position our contribution within the literature in \cref{sec:related} and
discuss final remarks and future development in \cref{sec:conclusion}.

%% file: implementation.tex
\section{Implementing a SAFARI Prototype}
\label{sec:implementation}

In this section, we present our prototypical implementation of SAFARI. We
outline its architecture and detail the concepts, technologies, and tools
employed for and their respective roles in the implementation. 

The fundamental concepts behind SAFARI's efficiency are: \emph{a})
\emph{Infrastructure as Code} (IaC), which is a paradigm where an orchestrator
manages the provisioning of infrastructure components, like computing, network,
and storage devices, using code rather than manual configuration,
\emph{b})\emph{OS-agnostic Task Automation} (OTA), which offers a uniform
interface that allows users to execute processes independently of the underlying
operating system, and \emph{c}) \emph{Investigation Tools}, which are the
software that provide visibility/metrics on malware's behaviour.

Our prototype reifies IaC with
Terraform\footnote{\url{https://www.terraform.io/}} and OTA with
Ansible\footnote{\url{https://www.redhat.com/en/ansible-collaborative}}. Since
our application context is crypto-ransomware, we concretise IT with ad-hoc
tools that generates a report of the files and checksums of a target VM and
compare the reports from before and after running tests to collect the
experiments' data.

Before delving into our prototype's architecture, we briefly describe which
hypervisor technology we choose. Indeed, while SAFARI's design abstracts away
from a specific virtualisation technology, one needs to fix it when considering
a specific deployment. We choose
Proxmox\footnote{\url{https://www.proxmox.com/}}, which is a widely-used and
renowned open-source type-2 hypervisor (hosted in Linux/Debian) that supports
enterprise-level virtualisation and includes an integrated web-based interface
that complements the API-based one for the management of virtual machines,
software-defined storage, and networking.

\subsection{Software Architecture}

\begin{figure*}[t]
    \centering
    \includegraphics[width=1\textwidth]{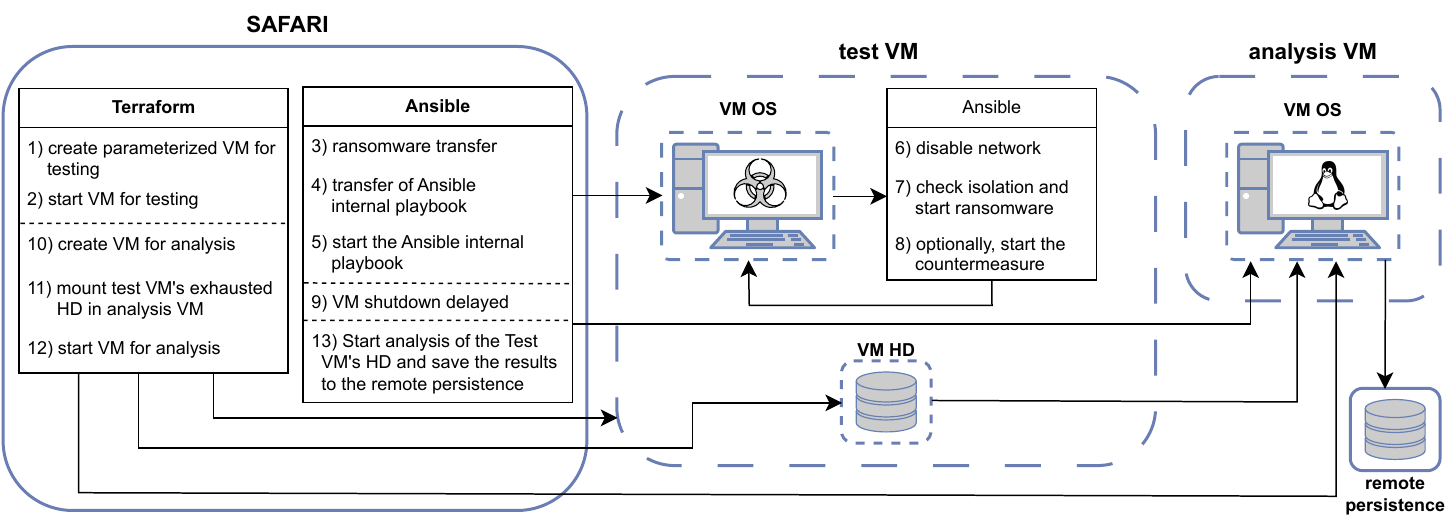}
    \caption{SAFARI's prototype architecture and functionalities.\label{fig:safari}}
\end{figure*}

We illustrate the architecture and experiments behaviour of our prototype in
\cref{fig:safari}. The main components of the architecture are the prototype
software components (labelled ``SAFARI'' in the figure, for brevity),
test VMs, analysis VMs, and remote persistence. We use borders, in
the figure, to represent ephemerality with a dashed line and persistence
with a solid one. The VMs that run the experiments (test VM) and the related
analysis (analysis VM) correspond to a test run and are ephemeral and discarded
within the related test session. The prototype's software and persistence
components respectively orchestrate the experiments and store their results (encompassing all the test sessions).

In \cref{fig:safari}, from the left, we find the core components of the
SAFARI prototype, which orchestrate the operations required to carry out the
tests. Terraform and Ansible scripts mainly conduct these operations. The
numbered operations shown in the schema correspond to a test session,
which one can run multiple times and in parallel to gather statistical
experimental data.

We describe the workflow of the prototype following the sequence in \cref{fig:safari}.

At the start of an experiment, the system interacts with the hypervisor to
create a VM (1) and to start it (2).
Then, control passes to OTA, which transfers the malware to the VM (3), along
with a list of tasks that the VM must execute (4). This step is peculiar and
particularly important for the \emph{air-gapping} principle mentioned in
\cref{sec:intro}. We require the VMs to run an internal OTA engine. In this way,
we avoid the network-based orchestration of the ransomware and rather inject a
part of the OTA tasks within the VM. We adopt a multi-tiered approach to enforce
isolation, which leads us to divide the logic of OTA orchestration into two
parts, one external and one internal. The internal one disables the network
connection (6) and makes sure that the VM is isolated before starting the
ransomware (7). The in-VM OTA component can also start possible countermeasures
present in the VM (8), according to the experiment's design.

All experiments have a set timeout, which triggers (9) the shutdown of the test
VM. Then, control passes back to IaC for the creation of the analysis
VM (10), which mounts the disk of the test VM for analysis and the remote
persistence disk for storing the analysis results (11, 12). OTA orchestrates
this step (13), which uses the IT tools to analyse the files and generate the results.

Since SAFARI simplifies managing VMs running different operating systems, our
recommendation is, e.g., to use Linux VMs for the analysis of Windows-specific
ransomware, to avoid possible accidental activations of the ransomware that
would threaten the reliability of the analysis.

\subsection{Filechecker and Profiler as IT}
We close this section with details of the other software contribution presented in this paper for the implementation of IT tools in SAFARI to study ransomware.

\begin{figure}[t]
    \centering
    \includegraphics[width=\columnwidth]{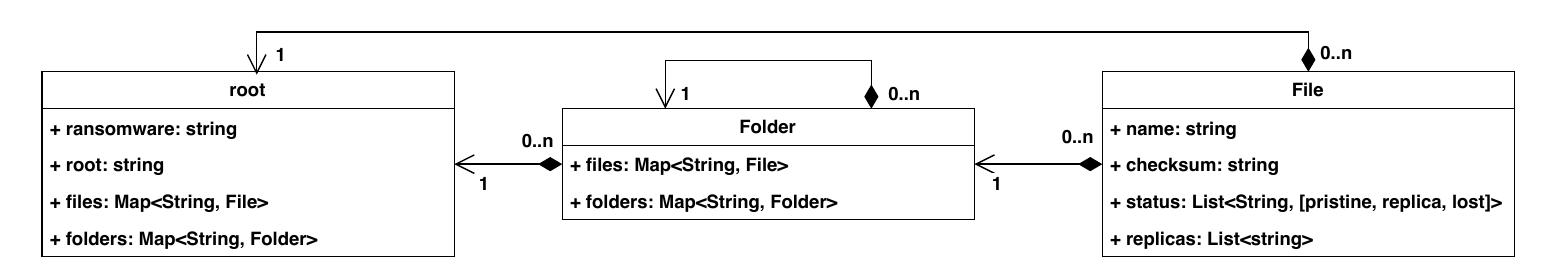}
    \caption{Schema of the Hierarchical Profile.}
    \label{fig:json-profiler}
\end{figure}

To implement the IT part of our SAFARI prototype, we develop a set of
open-source tools, available at
\url{https://github.com/Flooding-against-Ransomware/profiling}. We use these
tools to compare the state of VMs before and after a crypto-ransomware attack.
Specifically, we use a Filechecker tool to create VM file \emph{reports}. The
software generates a JSON-formatted list of the \emph{path}s of all the files
(descending within folders) contained within a given \emph{root} location,
associated with their \emph{checksum} signature (MD5). We expect investigators
to launch the Filechecker manually (one can automate this task with tools like
Vagrant\footnote{\url{https://github.com/hashicorp/vagrant}}.) when they
assemble a test VM template, so they generate a report of the pristine test VM
for later use in the analysis VM. Once obtained the report of the test VM, we
run a second tool, called Profiler. The Profiler, given two report files, a
reference one and a post-attack one, generates a JSON \emph{profile} file of the
difference between the two input files. Specifically, the profile indicates
which files are \emph{pristine}, which have been \emph{lost} due to encryption,
and possible \emph{replicas} that one can use to restore the original files of
the user (i.e., files with a different location but the same content of files of
the reference report). Interestingly, while having the pristine-lost ratio would
already give us a reading of the effectiveness/efficiency of ransomware and
contrast tools, measuring the replicas sheds further light on the modality of
execution of ransomware, since many of these create copies of the files to
encrypt before deleting them~\cite{KAMRK16}. In addition to a file-to-file
comparison, the Profiler generates a \emph{hierarchical} view of the profile,
following file locations, i.e., starting from the root, we find a list
\textit{files} contained therein, with their related status
(pristine/lost/replica) and a list of \textit{folders}, each containing a list
of files and subfolders. For reference, we report in \cref{fig:json-profiler}
the structure of hierarchical profiles---which also contain contextual
information of the experiment/analysis, like the name of the ransomware and the
root location (according to the reports).
The Profiler can process a hierarchical profile to generate a \emph{summary}
profile that indicates measures such as the total numbers of
pristine/lost/replica files found in hierarchical order---so that the root
reports the overall numbers, then broken down by the folders it contains---and
also aggregated by extension (useful to investigate e.g., which file formats a
given ransomware mainly targets).
Finally, the Profiler can \emph{aggregate} multiple summary profiles of the same
kind of experiment to provide statistical data about it. The result of is a JSON
profile with the same structure of the summary one but with additional elements
that report statistical measures on the files such as averages and standard deviations. 
By integrating the Filechecker and Profiler in our SAFARI prototype, we obtain
the automatic generation of statistics about the execution of ransomware and
their countermeasures.

%% file: case_study.tex
\section{Case Studies}
\label{sec:CaseStudies}

To illustrate the usage of SAFARI, we use our prototype to conduct two case
studies. The first regards the analysis of the behaviour of a set of known
ransomware, to profile their activity. The second showcases the integration
within tests of a ransomware countermeasure, Ranflood, to benchmark its
effectiveness against attacks. Before describing the case studies and their
results, we report on the deployment used to run an instance of our SAFARI
prototype.

\subsection{SAFARI's Prototype Deployment Infrastructure}

\begin{figure*}[t]
    \centering
    \includegraphics[width=\textwidth]{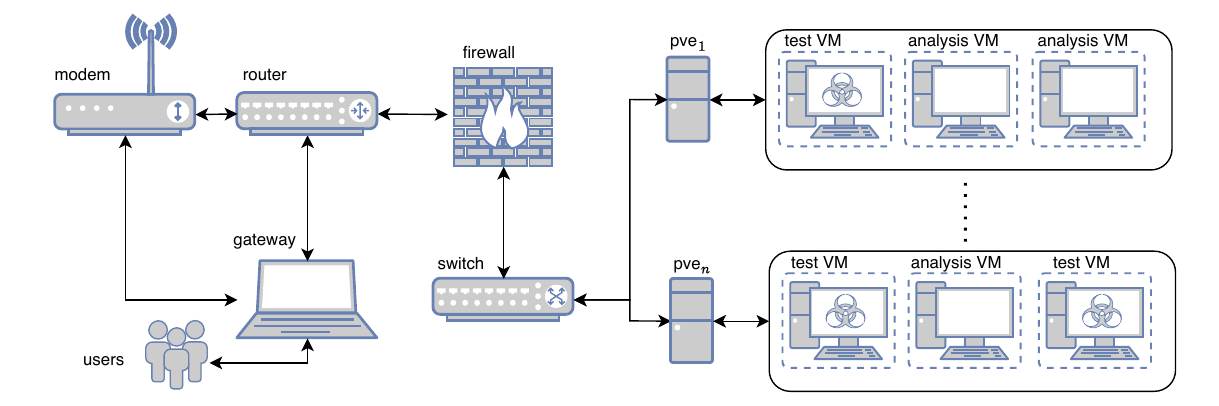}
    \caption{SAFARI's Prototype Deployment Infrastructure.}
    \label{fig:rig-architecturel}
\end{figure*}

The case studies helps us illustrate an on-premises deployment of our prototype.
We build the infrastructure shown in \cref{fig:rig-architecturel} as a companion
contribution to SAFARI's definition, designed to let users remotely run safe experiments.

Following \cref{fig:rig-architecturel}, users can authenticate and execute the
prototype's functionalities through a gateway connected to the Internet. From
the gateway, users can interact with the nodes, named pve\(_1\), \dots,
pve\(_n\) (where pve stands for Proxmox virtualisation environment), that make
up the cluster and host the test and analysis VMs. All these nodes and VMs are
behind a firewall that regulates access to the Internet through the router and
via the modem.

We detail the main features of the components found in
\cref{fig:rig-architecturel}. The \emph{gateway} runs Debian v.12 (although
unlikely, we use a Linux machine to further stave off possible infections from
Windows ransomware) and represents the only network access point that
experimenters use to connect to the nodes. Users connect to the gateway via the ZeroTier VPN and use sshuttle\footnote{Resp.\@ \url{https://www.zerotier.com/}
and \url{https://github.com/sshuttle/sshuttle}.} to access directly the nodes,
i.e., to run commands as if executed on machines in their
local network.

The \emph{router}, mounting OpenWrt~\cite{websiteOpenWrt}, can give Internet
connection to the nodes in the cluster, but by default its \emph{firewall}
prevents VMs from accessing the Internet to avoid the possible propagation of
ransomware.

The hardware of the pve cluster represents typical office/desktop
personal computers, and it encompasses eight machines.
The nodes run ProxMox 7.0-8 and include two VM templates, Linux Ubuntu
24.04 for the analysis VM and a Windows 10 (x64) 1809 template for the
test VM. While each node can run multiple VMs, to have good performance, we run
at most two VMs per node.

\subsection{Case Study: Ransomware Profiling}
\label{sec:case-study}

\begin{figure}[t]
\centering
\includegraphics[width=0.28\columnwidth]{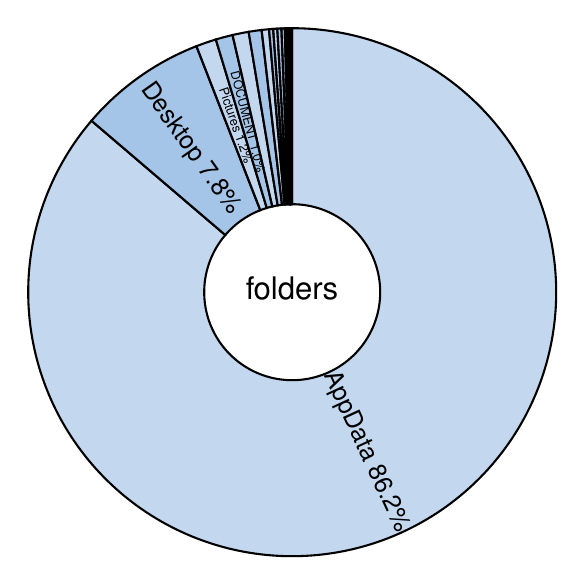}
\includegraphics[width=0.28\columnwidth]{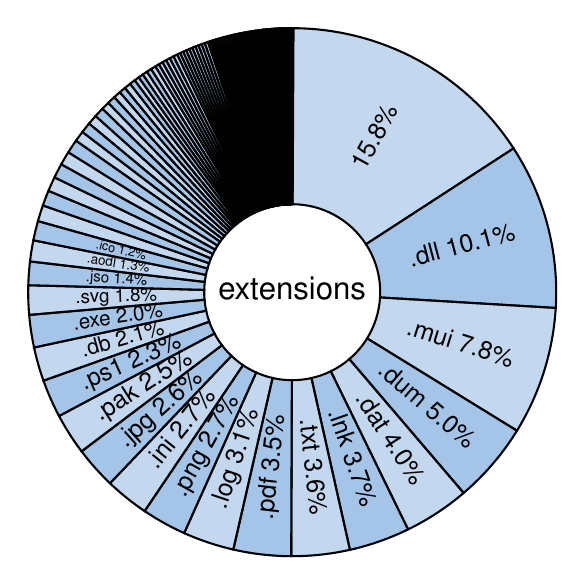}
\caption{Visualisation of the user's file profiles.\vspace{-1em}}
\label{fig:profiling_ransomware_base}
\end{figure}

To conduct the testing of our prototype, we start by defining a testing
protocol, comprising various types of ransomware. We base the selection of
these ransomware samples on an analysis of the history of ransomware and its
evolution over time~\cite{10.1145/3514229}.
We source the samples used for the tests from repositories
previously used and recognised in other academic contexts~\cite{article},
such as VirusTotal.

\begin{figure*}[t]
    {\setlength{\tabcolsep}{0pt}
    \begin{tabular}{c|c|c|c|c}
    & Summary & Pristine & Replica & Lost \\
    \hline
    \hline
    \rotatebox{90}{\adjustbox{width=5em}{Vipasana (Ryuk)}}\hspace{.25em} &
    \includegraphics[width=0.135\textwidth]{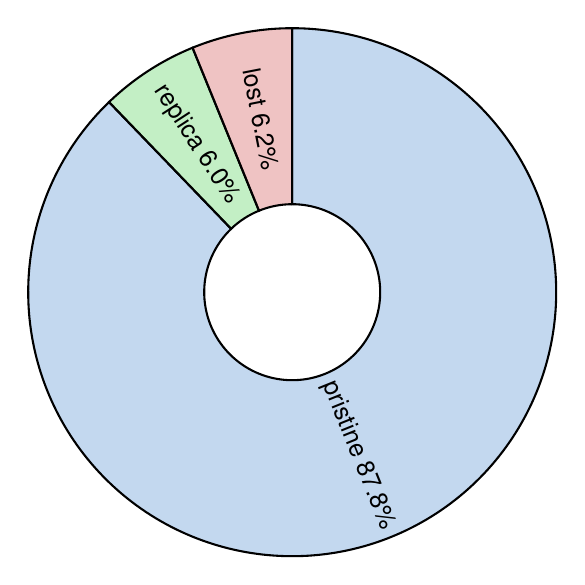}   
    &
    \includegraphics[width=0.135\textwidth]{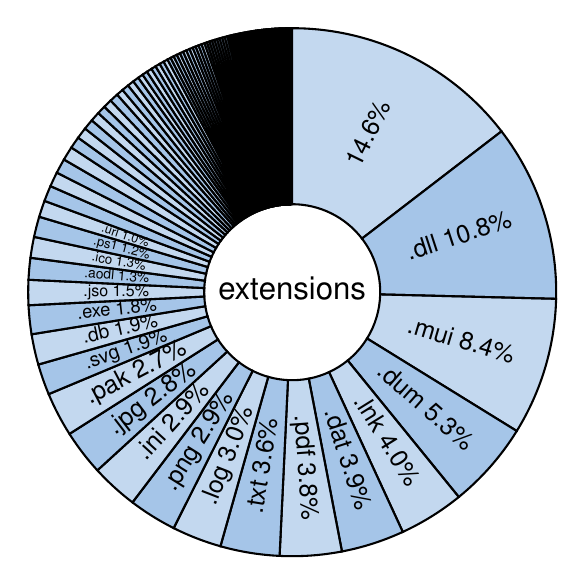}   
    \includegraphics[width=0.135\textwidth]{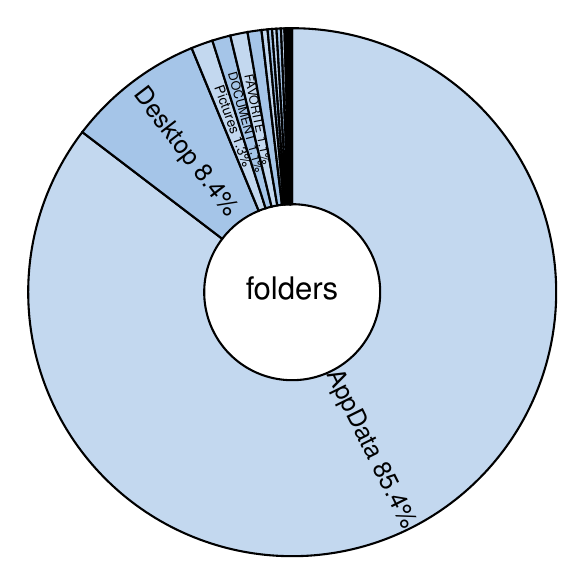}    
    &
    \includegraphics[width=0.135\textwidth]{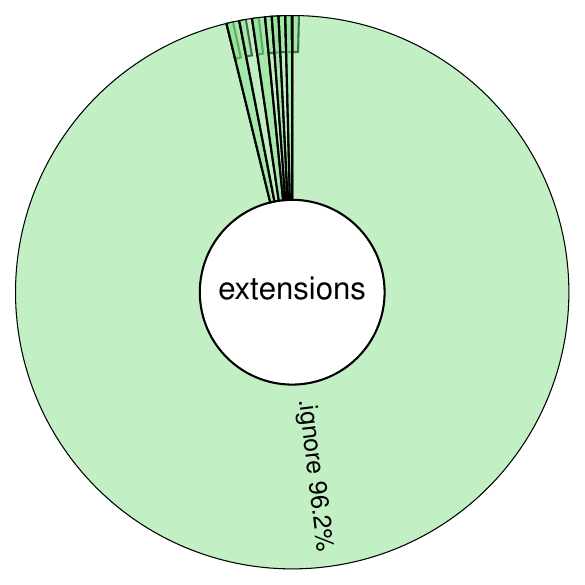}    
    \includegraphics[width=0.135\textwidth]{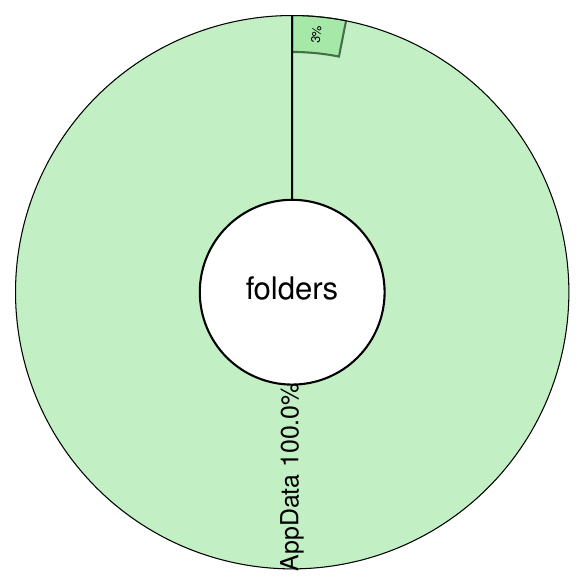}     
    &  
    \includegraphics[width=0.135\textwidth]{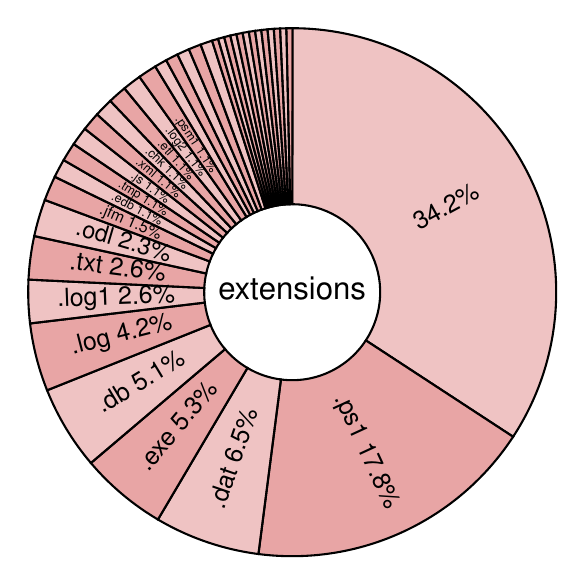}
    \includegraphics[width=0.135\textwidth]{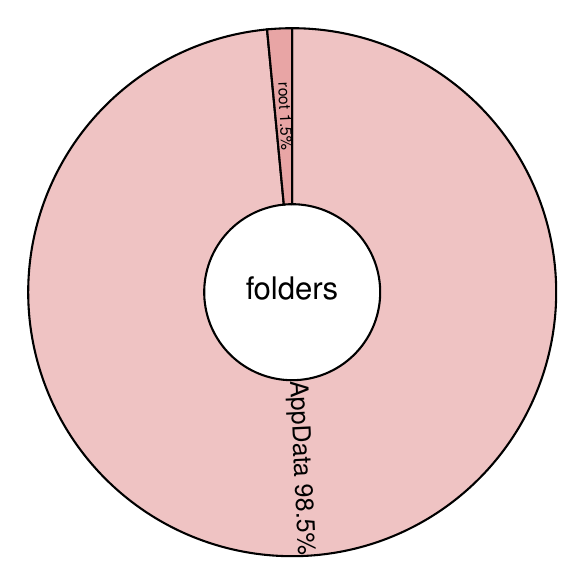}
    \\
    \hline
    {\rotatebox{90}{WannaCry}}\hspace{.25em} &
    \includegraphics[width=0.135\textwidth]{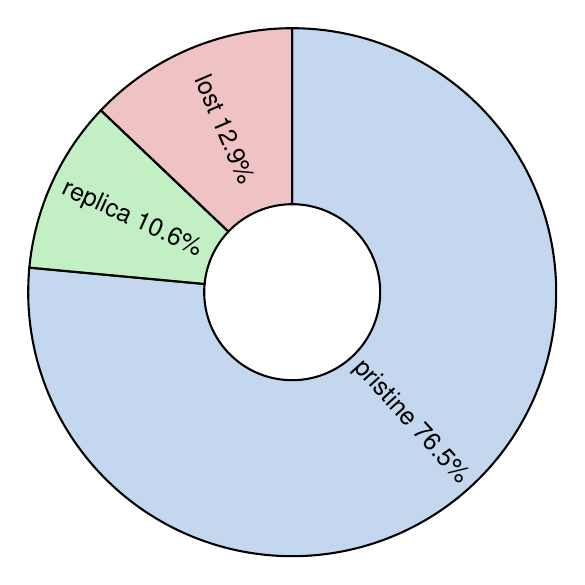}   
    &
    \includegraphics[width=0.135\textwidth]{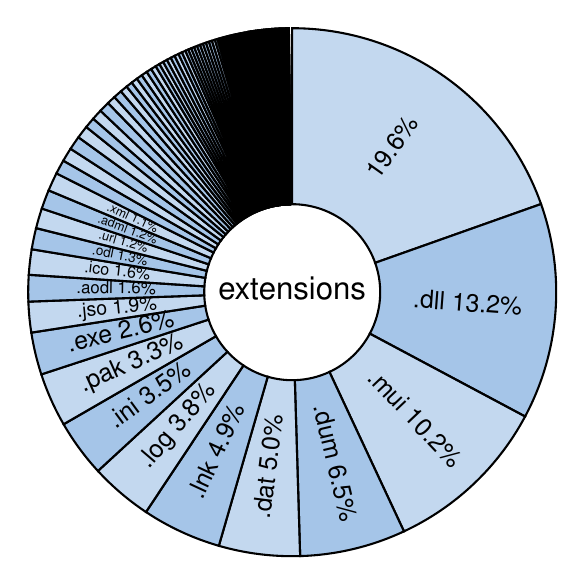}   
    \includegraphics[width=0.135\textwidth]{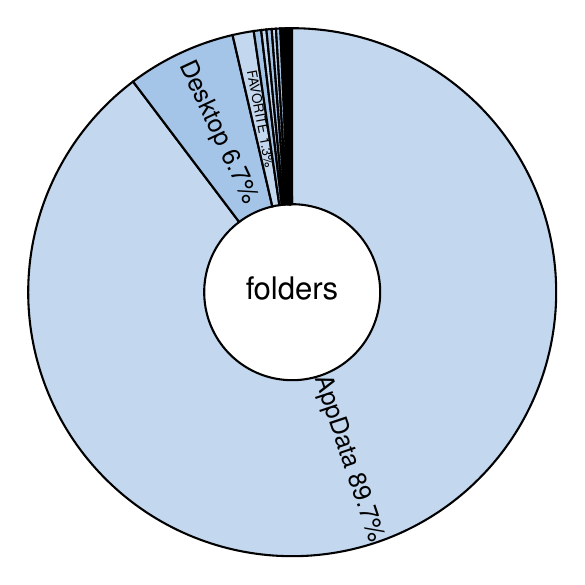}    
    &
    \includegraphics[width=0.135\textwidth]{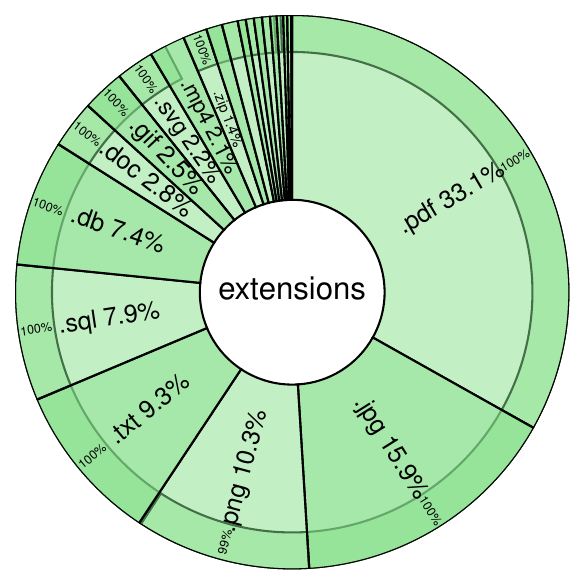}    
    \includegraphics[width=0.135\textwidth]{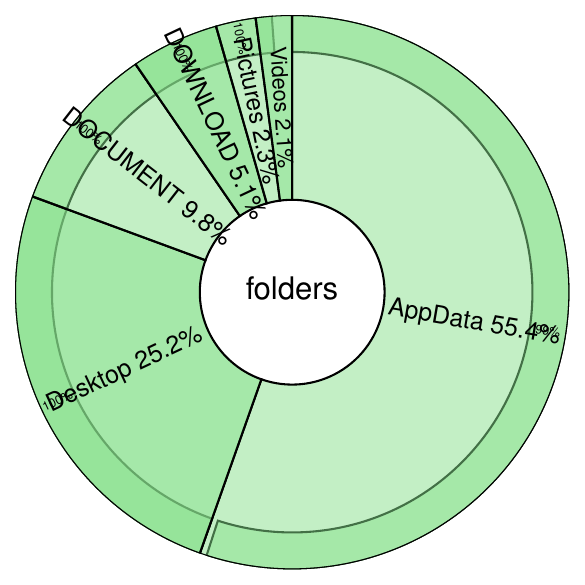}     
    &  
    \includegraphics[width=0.135\textwidth]{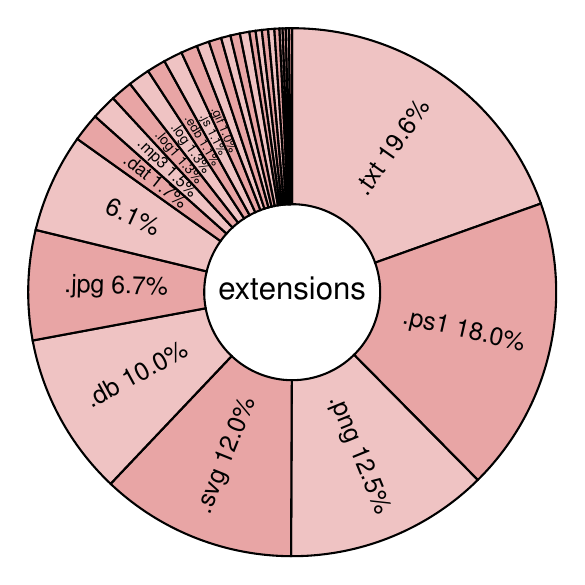}
    \includegraphics[width=0.135\textwidth]{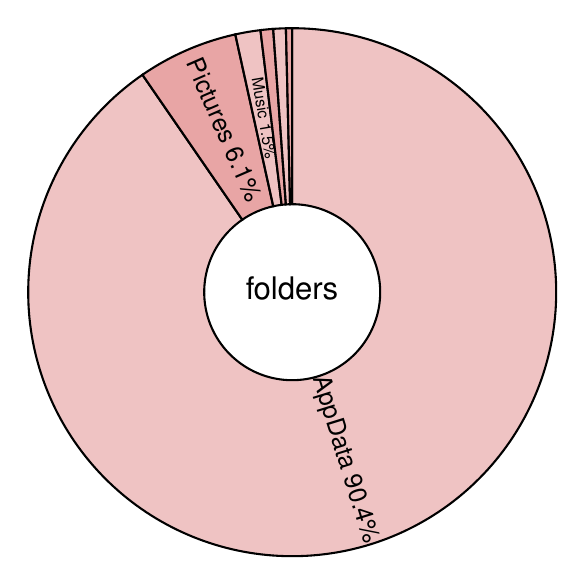}
    \\
    \hline 
    \raisebox{.5em}{\rotatebox{90}{LockBit}}\hspace{.25em} &
    \includegraphics[width=0.135\textwidth]{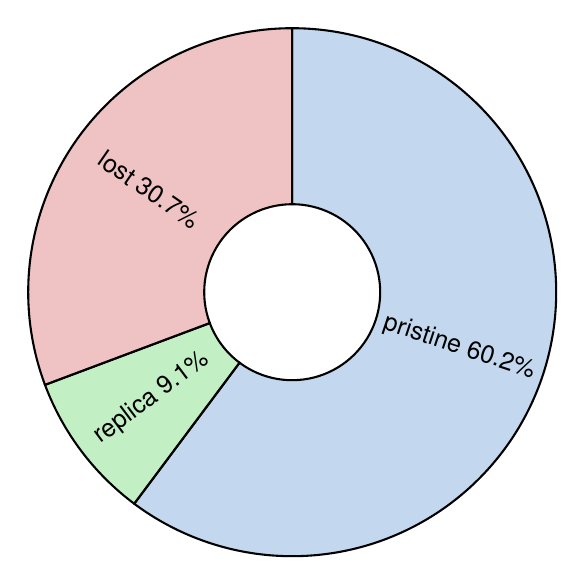}   
    &
    \includegraphics[width=0.135\textwidth]{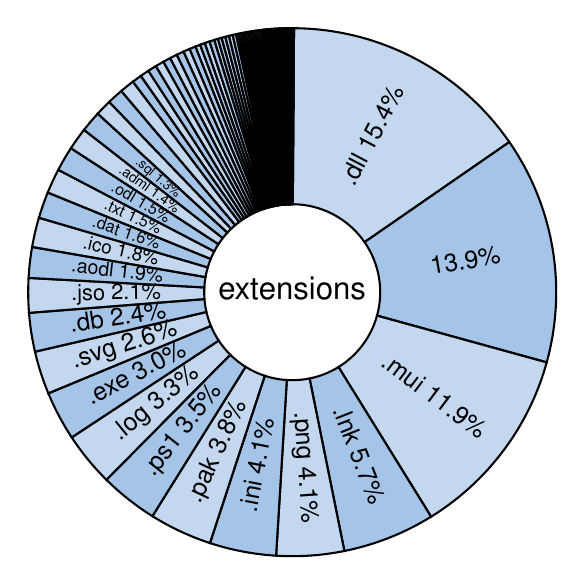}   
    \includegraphics[width=0.135\textwidth]{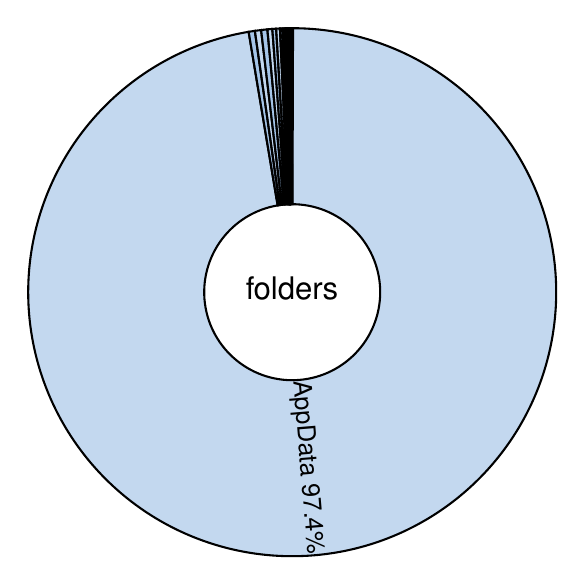}    
    &
    \includegraphics[width=0.135\textwidth]{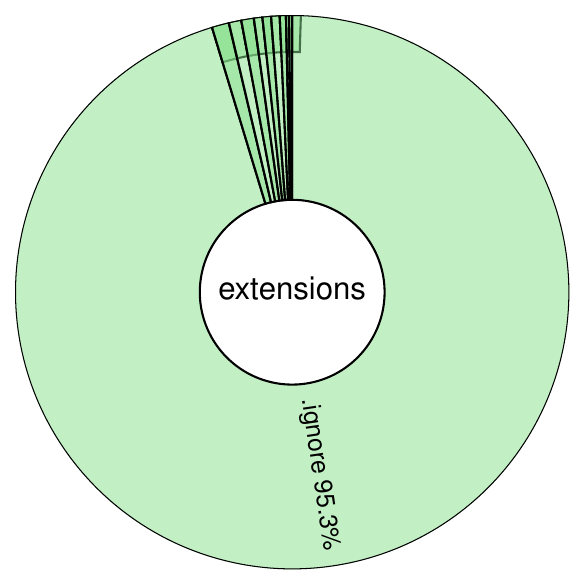}    
    \includegraphics[width=0.135\textwidth]{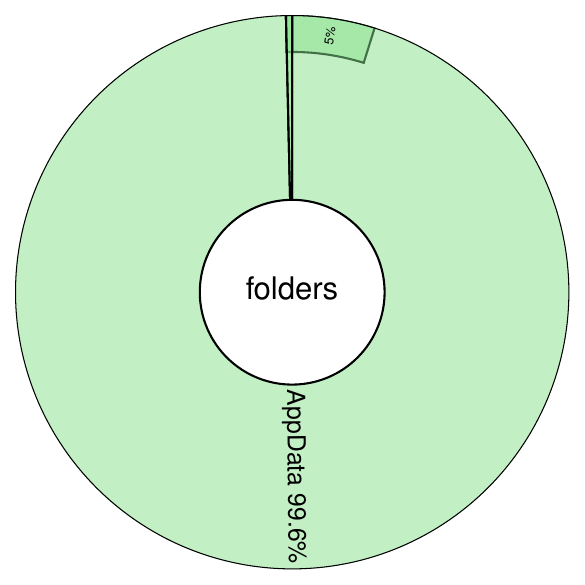}     
    &  
    \includegraphics[width=0.135\textwidth]{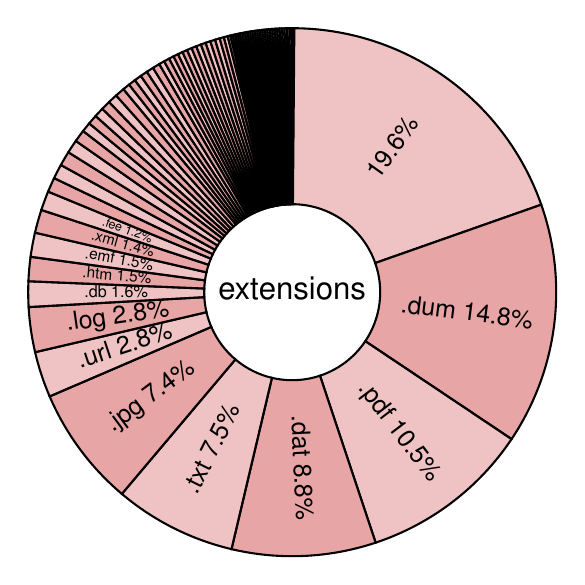}
    \includegraphics[width=0.135\textwidth]{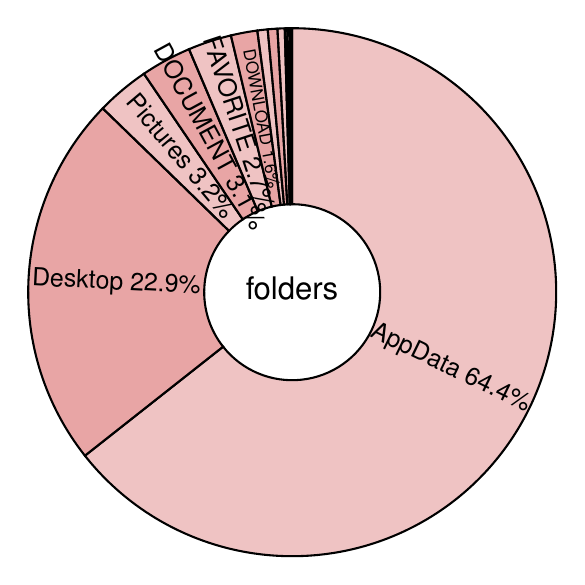}
    \\
    \hline
    \raisebox{.5em}{\rotatebox{90}{Phobos}}\hspace{.25em} &
    \includegraphics[width=0.135\textwidth]{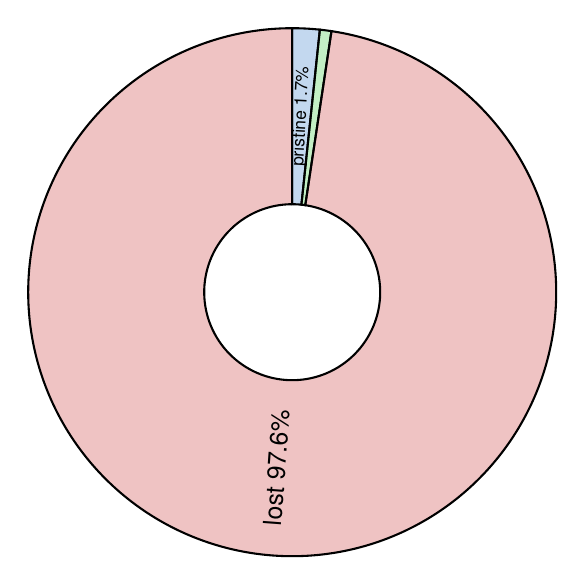}   
    &
    \includegraphics[width=0.135\textwidth]{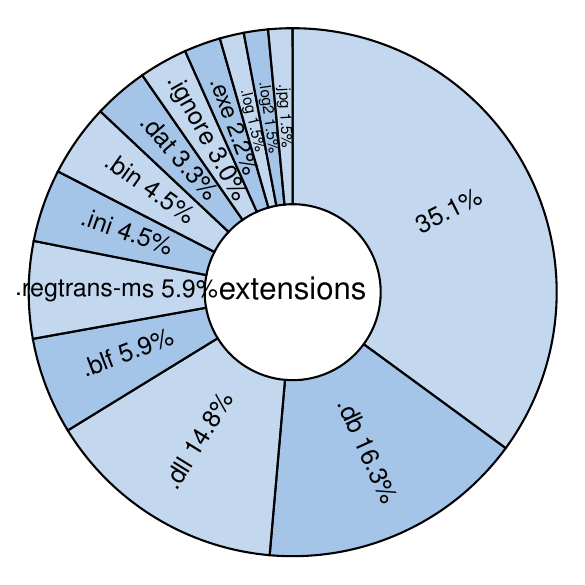}   
    \includegraphics[width=0.135\textwidth]{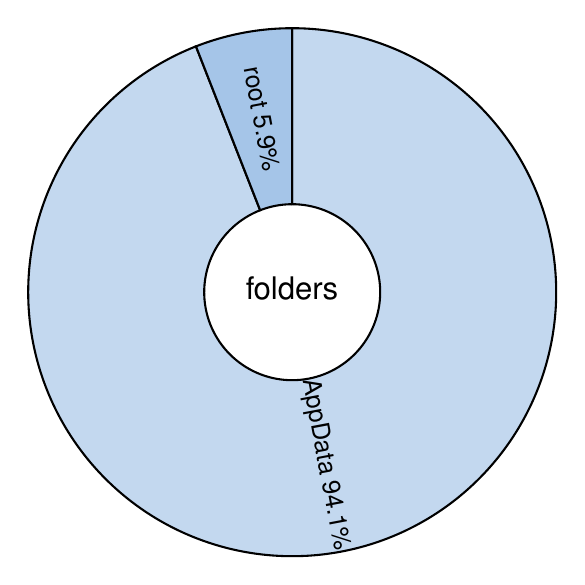}    
    &
    \includegraphics[width=0.135\textwidth]{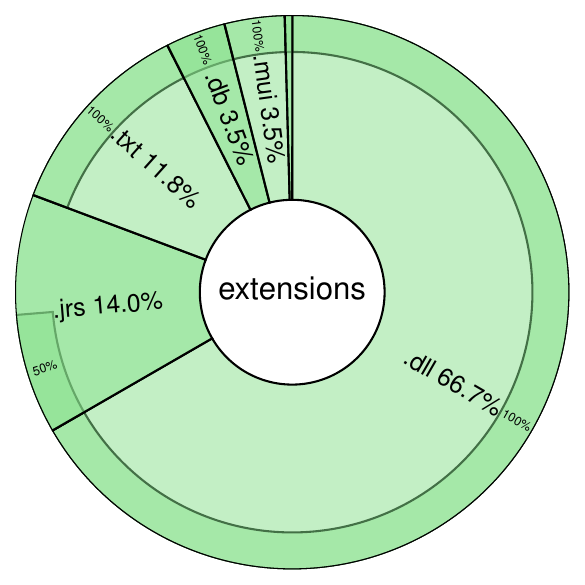}    
    \includegraphics[width=0.135\textwidth]{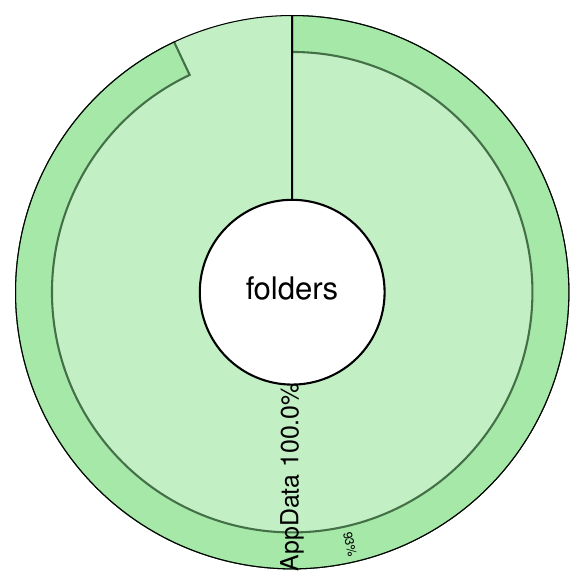}     
    &  
    \includegraphics[width=0.135\textwidth]{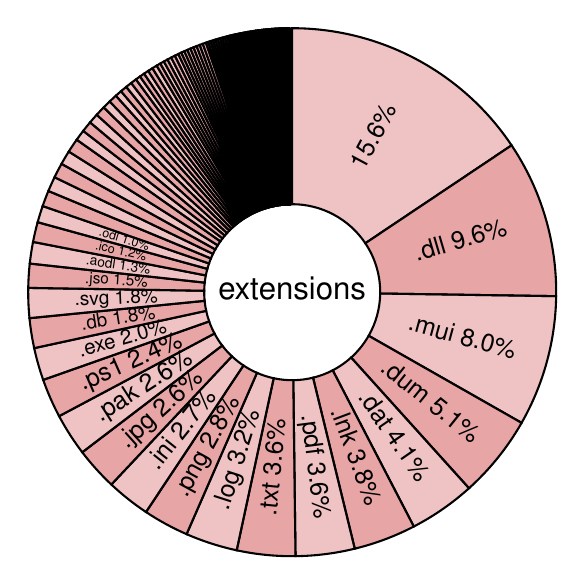}
    \includegraphics[width=0.135\textwidth]{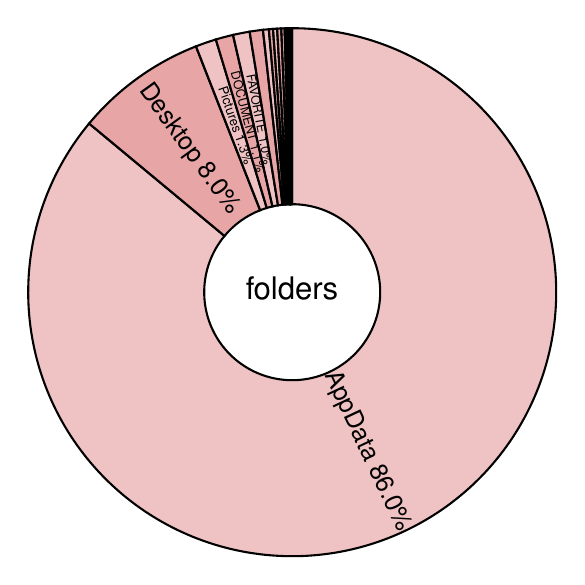}
    \\
    \end{tabular}}
    \caption{Attack profiles of Vipasana/Ryuk, WannaCry, LockBit, and Phobos.}
    \label{fig:profiling_ransomware}
\end{figure*}

In the first case study, we demonstrate our SAFARI prototype's capability to analyse the encryption patterns of five well-known Windows ransomware: Ryuk, Vipasana, WannaCry, LockBit, and Phobos. To simulate a real-world scenario, we configured the test VM with a typical user profile, modelled after Hansley~\cite{Halsey2016}, including the primary directories of the Windows 10 file system. The profile features 2 GB of user data, aligned with file type recommendations from Scaife et al.~\cite{7536529} and Kaspersky~\cite{Kasp}, along with file names commonly targeted by ransomware~\cite{Kroll}. Additionally, it incorporates user-interactive programs such as a web browser and an office suite~\cite{RDGKPPBV12}. The profile includes 13 ``user'' folders, such as ``Documents'', ``Desktop'', ``Music'', and ``Pictures'' with the breakdown reported in \cref{fig:profiling_ransomware_base}, using the prototype's Filechecker and Profiler tools.

For each ransomware strain, we run 32 experiments with a timeout of 10 minutes
for each experiment. We obtain at least 16 valid experiments for each ransomware
--- 22 for LockBit, 32 for Ryuk and Vipasana, and 16 for Phobos and Wannacy,
where we discard the ones where the ransomware did not start.

We visualise the results in \cref{fig:profiling_ransomware}, reporting in the
rows the data of a given ransomware and divide these (from left to right)
respectively between the percentage summary of pristine/lost/replica files and
the percentage breakdown of pristine, replica, and lost files by extensions and
folders. As Ryuk and Vipasana exhibit similar behavior, only Vipasana's results
are shown for brevity. Replica file plots include smaller outer segments
indicating the percentage of distinct replicas within a partition, e.g., the 3\%
in Ryuk/Vipasana folder replicas denotes that only 3\% are unique, with the
remainder having multiple copies.

Before analysing the results, we observe that all ransomware generate replicas,
leading to two key insights. First, the samples employ a copy-based strategy,
involving a copy phase followed by encryption --- the replicas found are
temporary files left unencrypted due to the VM shutdown. Second, the presence of
replicas is a performance indicator, suggesting that the 10-minute delay was
insufficient for the ransomware to encrypt all target files.

We comment on the results in \cref{fig:profiling_ransomware} from top to bottom.

Ryuk and Hiragana exhibit similar behaviour, primarily targeting the ``AppData'' folder (containing application files restorable through reinstallation). They create ``.ignore'' files in ``AppData'', often making multiple copies (3\% are distinct). This strategy likely aims to prevent recovery by simple renaming.

WannaCry also focuses on ``AppData'' but spreads its attack across other directories like ``Pictures'', ``Music'', ``Desktop'', and ``Documents''. It creates unique copies before encryption and targets common file types such as ``.png'', ``.jpg'', ``.svg'', ``.pdf'', ``.txt'', and ``.doc'' while skipping files without extensions.

LockBit outperforms WannaCry, encrypting 30\% of files compared to WannaCry's 12\%. It targets user directories like ``Desktop'' (23\%), ``Pictures'', and ``Documents'' while still focusing on ``AppData''. Like Ryuk/Vipasana, LockBit generates ``.ignore'' copies but also appears to use a three-stage strategy involving direct copying, renaming to ``.ignore'', and encryption.
Phobos seems the most dangerous one, reaching a staggering 98\% of lost files.
Given the almost-complete coverage of the attack area, looking at the lost and
pristine charts gives little insight on its behaviour, although we might infer
that its encryption routine might skip specific locations, found within the user
profile and ``AppData'' folders that mainly contain executable and configuration
files, e.g., preserved by the ransomware to allow minimal system functioning to
show the ransom message.

Summarising the results, we notice that the severity of the different types of
ransomware is not directly related to the mere percentage of lost files (which,
of course, is an important measure, per se). For example, Phobos has the highest
percentage of lost files while LockBit seems less dangerous since it encrypt
only ca.\@ 30\% of the user's files. However, the breakdown of the encrypted
files tells a more nuanced story. Phobos is much more ``blunt'' in its
behaviour, since it indiscriminately encrypts most of the user files,
irrespective of the folder and the extension. On the contrary, LockBit, is much
more efficient, since it focuses its effort only on the files that mainly matter
to the user, i.e., the files found under ``Desktop'', whose sequester compel the
user to pay the ransom.

\subsection{Ransomware Mitigation Profiling}
The second case study provides an example of how one can use SAFARI to evaluate
mitigation tools and techniques in a realistic yet safe scenario.

Concretely, we profile a recent ransomware tool, called Ranflood~\cite{BGMMP24},
exponent of a family of solutions, called Data Flooding against
Ransomware~\cite{BGMMOP23}. 
Essentially, Ranflood contrasts ransomware attacks by confounding the files of
the user with a ``flood'' of decoy ones, stymieing the attack both file- and
resource-wise (by contending IO access with the ransomware).

Ranflood provides two families of flooding strategies: \textit{random} and
\textit{copy-based}. The \textit{random} family, implemented by the
\textit{Random} strategy, generates files of varying sizes and formats
(mimicking those targeted by ransomware) with random content, requiring only a
disk location to flood.

The \textit{copy-based} family, implemented by the \textit{On-the-fly} and
\textit{Shadow} strategies, require a target location and a preliminary setup
under normal conditions: \textit{On-the-fly} collects checksums of pristine
files to avoid copying corrupted ones, while \textit{Shadow} creates backups of
user files to use as the source for flood copies.

\begin{figure*}[t]
    {\setlength{\tabcolsep}{0pt}
    \begin{tabular}{c|c|c|c|c}
    & Summary & Pristine & Replica & Lost \\
    \hline
    \footnotesize 
    {\rotatebox{90}{\adjustbox{width=5em}{WannaCry (Shadow)}}}\hspace{.25em} &
    \includegraphics[width=0.135\textwidth]{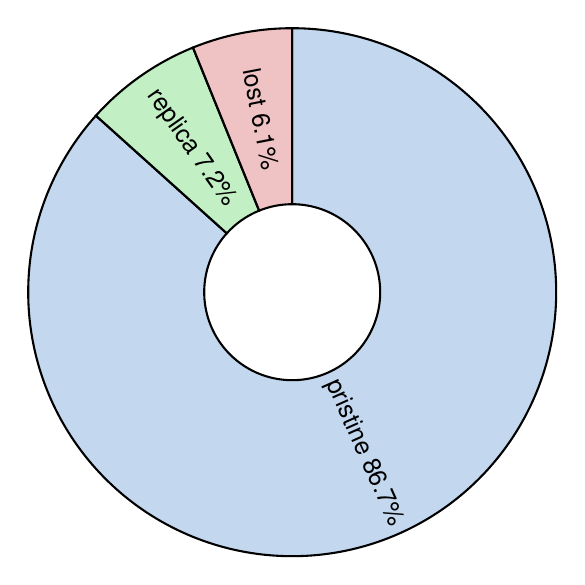}   
    &
    \includegraphics[width=0.135\textwidth]{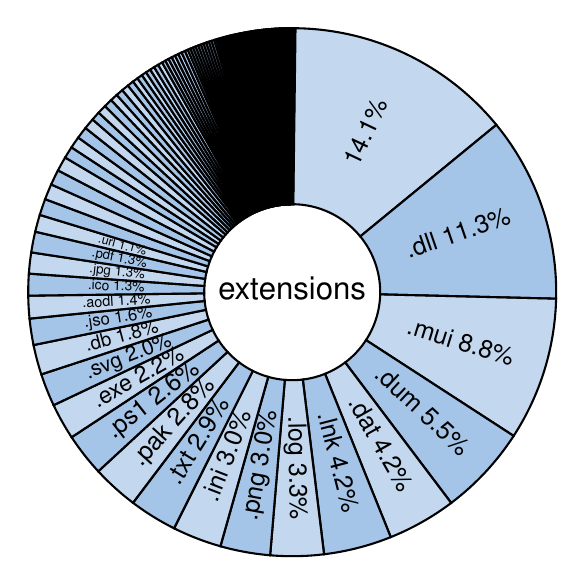}   
    \includegraphics[width=0.135\textwidth]{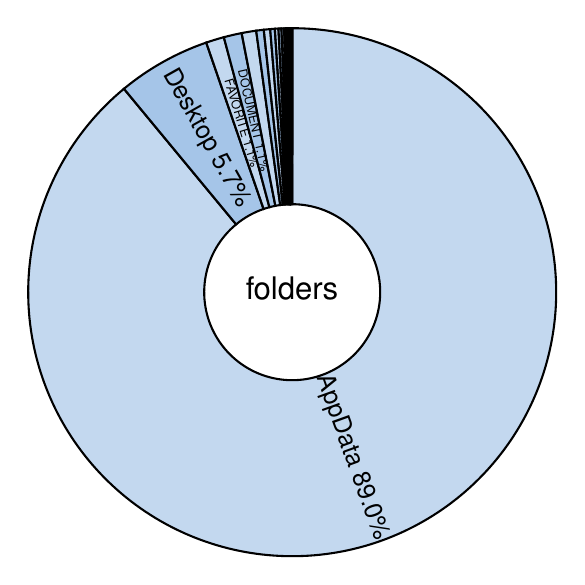}    
    &
    \includegraphics[width=0.135\textwidth]{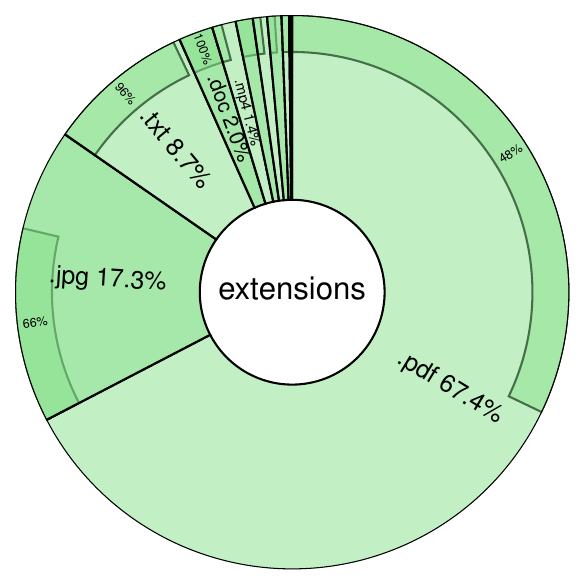}    
    \includegraphics[width=0.135\textwidth]{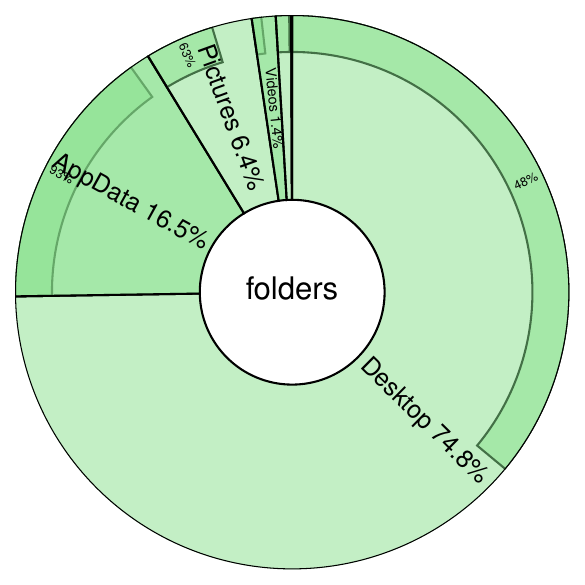}     
    &  
    \includegraphics[width=0.135\textwidth]{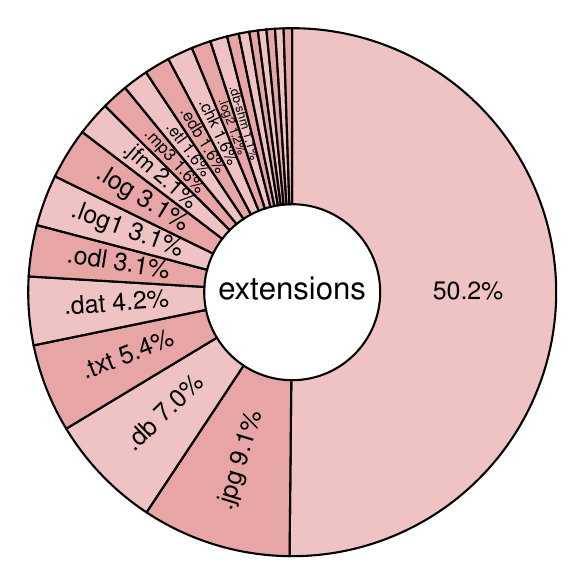}
    \includegraphics[width=0.135\textwidth]{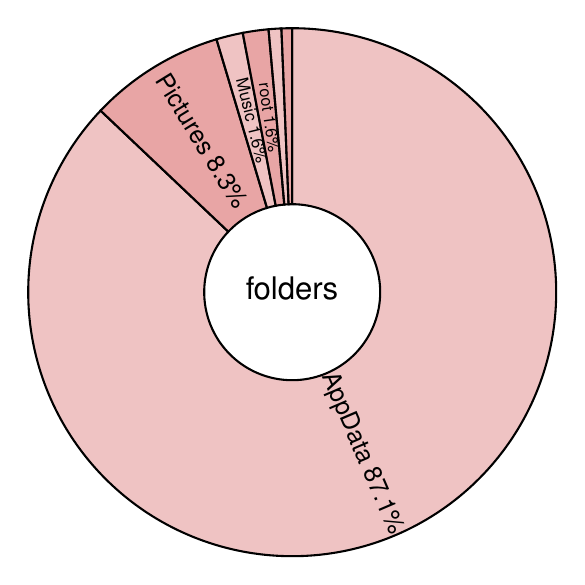}
    \\
    \hline
    \footnotesize 
    {\rotatebox{90}{\adjustbox{width=5em}{LockBit (Shadow)}}}\hspace{.25em} &
    \includegraphics[width=0.135\textwidth]{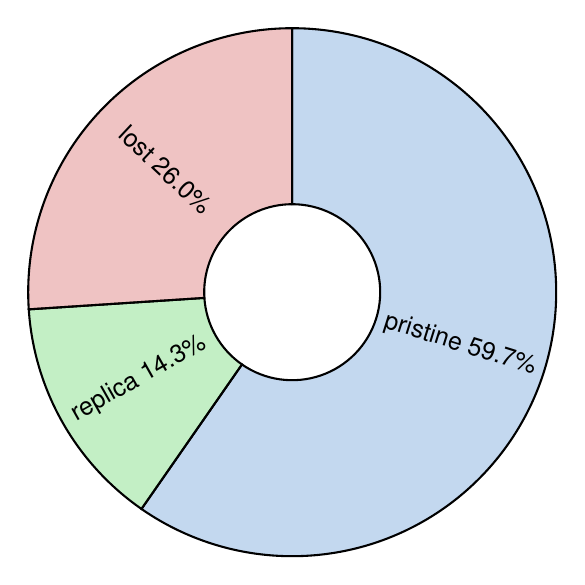}   
    &
    \includegraphics[width=0.135\textwidth]{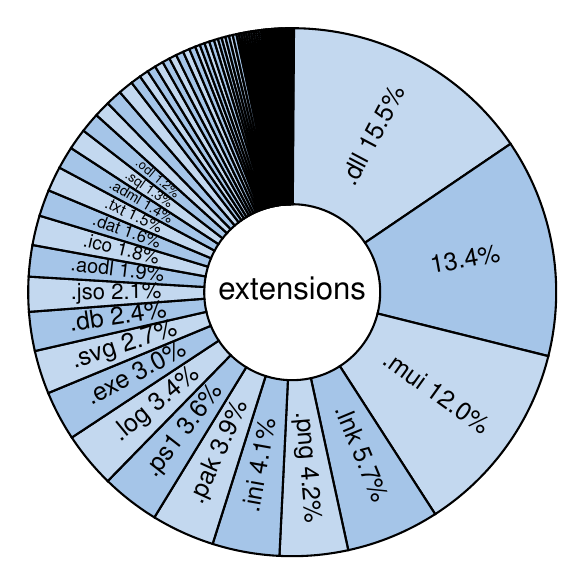}   
    \includegraphics[width=0.135\textwidth]{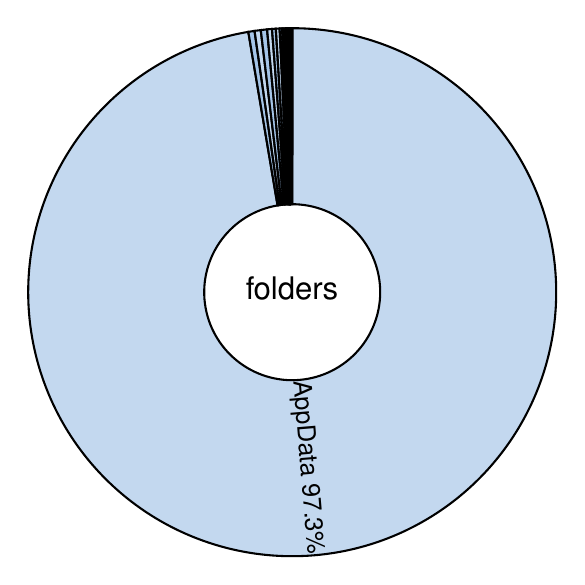}    
    &
    \includegraphics[width=0.135\textwidth]{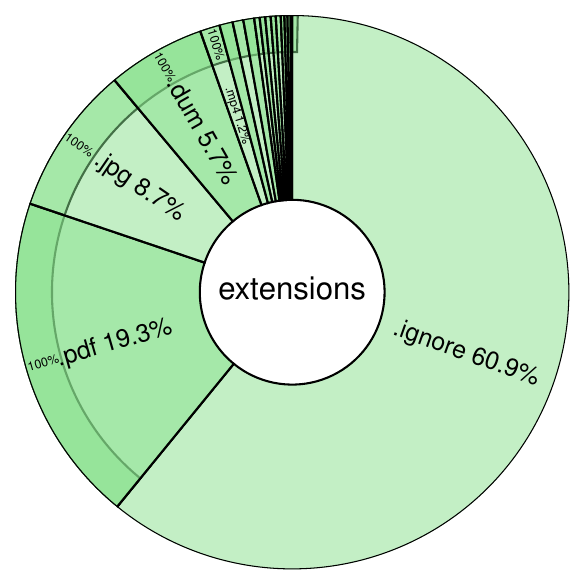}    
    \includegraphics[width=0.135\textwidth]{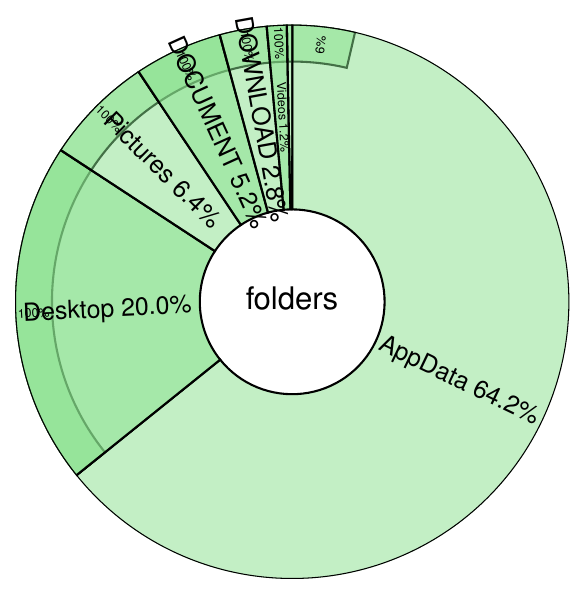}     
    &  
    \includegraphics[width=0.135\textwidth]{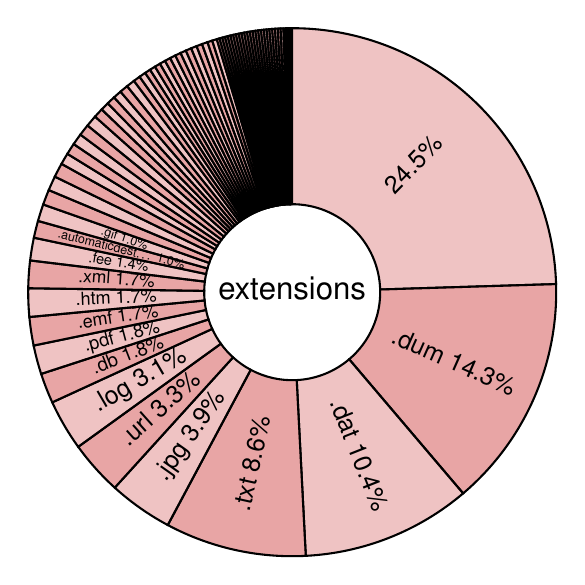}
    \includegraphics[width=0.135\textwidth]{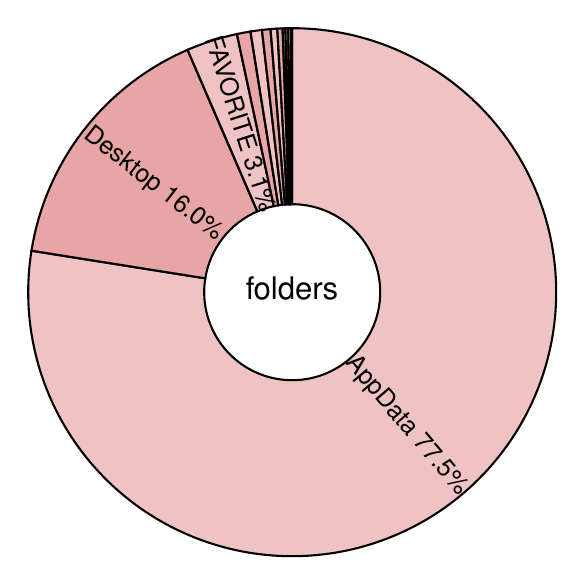}
    \\
    \hline 
    \footnotesize 
    {\rotatebox{90}{\adjustbox{width=5em}{Phobos (Shadow)}}}\hspace{.25em} &
    \includegraphics[width=0.135\textwidth]{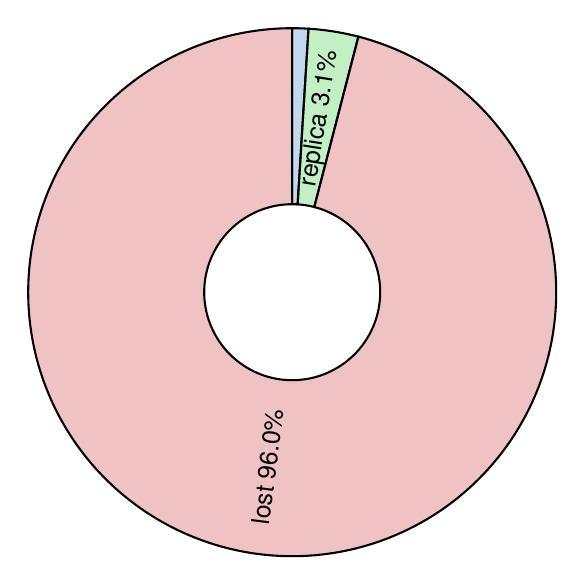}   
    &
    \includegraphics[width=0.135\textwidth]{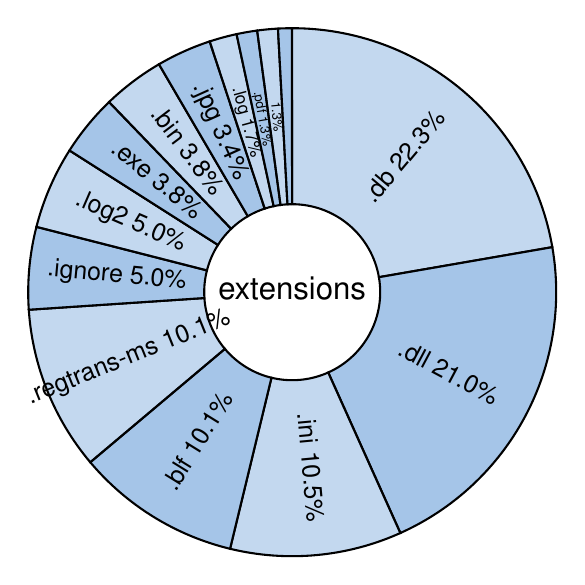}   
    \includegraphics[width=0.135\textwidth]{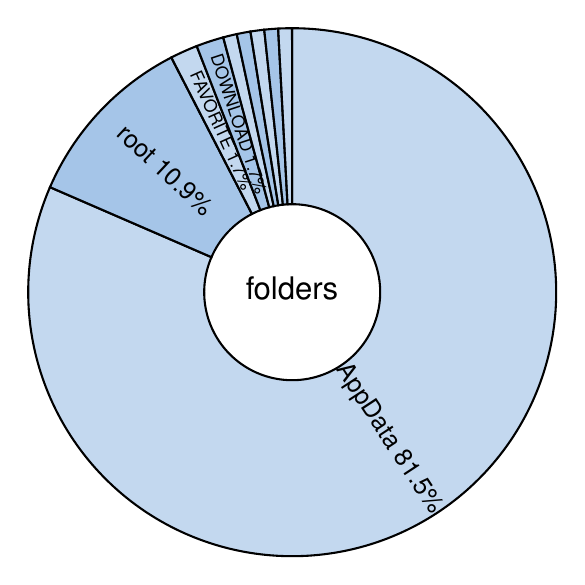}    
    &
    \includegraphics[width=0.135\textwidth]{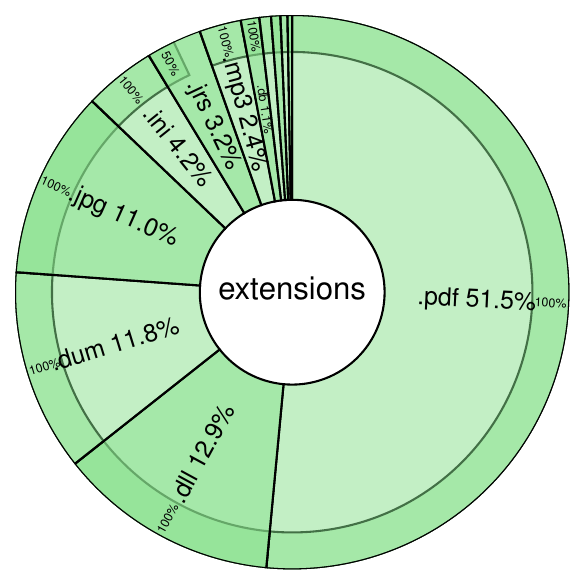}    
    \includegraphics[width=0.135\textwidth]{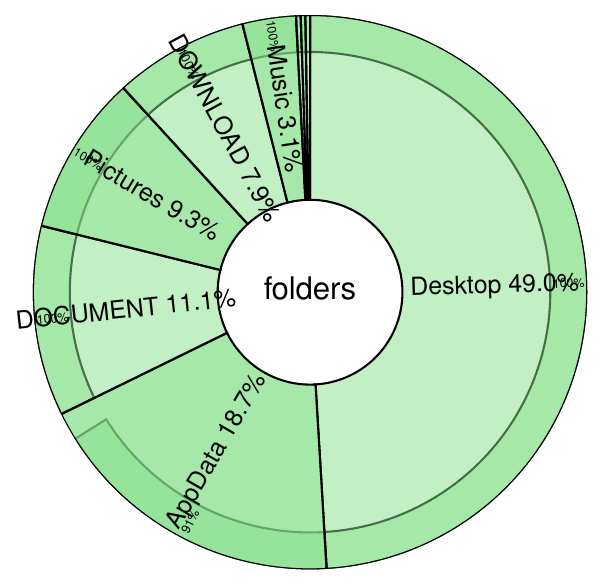}     
    &  
    \includegraphics[width=0.135\textwidth]{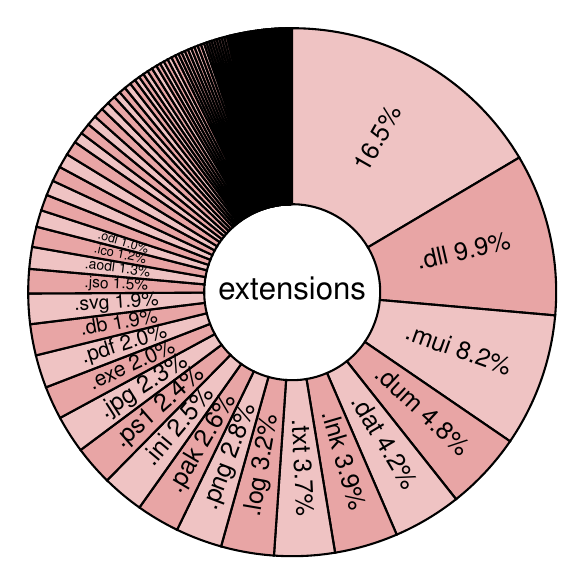}
    \includegraphics[width=0.135\textwidth]{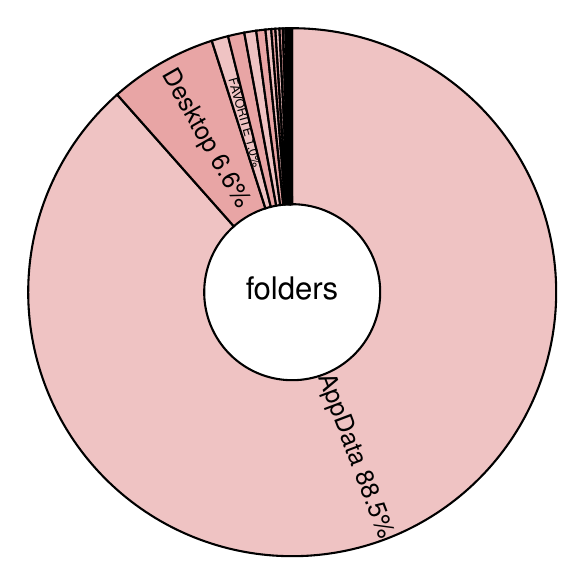}
    \\
    \end{tabular}}
    \caption{Ranflood's contrast profiles against WannaCry, LockBit, and Phobos.}
    \label{fig:profiling_ranflood}
\end{figure*}

Since Vipasana/Ryuk do not attack the files of the user
(cf.~\cref{fig:profiling_ransomware}), we concentrate on WannaCry, LockBit, and
Phobos. We report, in \cref{fig:profiling_ranflood}, the results of the
experiments after running 12 attack-and-contrast instances on the template VM
used in the first case study.
For each selected ransomware, we run the different Ranflood strategies with a
30-second delay since starting the ransomware, launching 13 flooding instances
in parallel on distinct folders of the user, including ``Documents'',
``Desktop'', ``Music'', and ``Pictures'', purposefully avoiding flooding the
``AppData'' folder which contains application files rather than the user's ones
and stopping the VM after 10 minutes.

For brevity, we show, in \cref{fig:profiling_ranflood}, only the best-performing
strategy, i.e., the one that minimises the percentage of lost files, which is
Shadow. The interested reader can find all data and plots at
\url{https://zenodo.org/records/13891513}. The data in the plots is the average
of 12 runs, amounting to an average standard deviation of 20.66\% (across
pristine, lost, and replica files).

Considering the specific ransomware, we find that Ranflood reduces the number of
lost files in all cases, albeit with different impacts---WannaCry -52\%, LockBit
-15\%, and Phobos -1.6\%. As expected, since the Shadow strategy is copy-based,
we find an increased number of replicas in most cases---LockBit +46.1\% and
Phobos +442.8\%---except for WannaCry, where we observe a decrease of 32\%.
While the latter result deserves further investigation, we conjecture it might
derive from the high number of replicas used by WannaCry during attacks (the
largest among the tested ransomware, cf.~\cref{fig:profiling_ransomware}), which
Ranflood hindered via IO access contention. The breakdown of the WannaCry
replicas in \cref{fig:profiling_ranflood} sheds some light on the phenomenon.
While, in \cref{fig:profiling_ransomware}, we find most (55.4\%) replicas under
the ``AppData'' folder and fewer (25.2\%) in ``Desktop'', the scenario with
Ranflood reverses the ratios, with most replicas under ``Desktop'' (74.8\%) and
fewer under ``AppData'', which we attribute to the flooding performed by
Ranflood that hinders the attacks of the ransomware.

The replica plots of LockBit and Phobos greatly differ from the respective ones
in \cref{fig:profiling_ransomware}, hinting at the ``fight'' put up by
Ranflood against the ransomware on folders like ``Desktop'' and ``Pictures''.

%% file: related_work.tex
\section{Related work}
\label{sec:related}

Since SAFARI simulates real-world scenarios of hosts in an emulated air-gapped
network, we can see it as an implementation of a Digital Twin (DT); a
transformative technology that predicts system failures and
identifies anomalies. 

SAFARI fits into such a technological paradigm, particularly Cyber Ranges. A Cyber Range~\cite{10.1145/3387940.3392199}, when viewed as a DT
application, is a virtualised environment designed to replicate real-world IT
systems, networks, and infrastructures for cybersecurity testing and
research~\cite{9847127}, without impacting live operations.

In the literature, practical technological implementations of this type are
scarce, or are developed around a vertical use-case scenario.

De Benedictis et al.~\cite{10049521} proposed an IIoT anomaly detection
architecture based on DT and autonomic computing paradigms, utilizing the MAPE-K
feedback loop to monitor, analyze, plan, and execute reconfiguration or
mitigation strategies based on deviations from prescriptive behavior stored as
shared knowledge. While effective for anomaly detection, it is limited to this
scope and cannot test malicious software like ransomware due to the absence of
air-gapping.

Another related work is the one presented by Masi et
al.~\cite{masi2023securing}, who derive a cybersecurity DT as part of the
security-by-design practice for Industrial Automation and Control Systems used
in Critical Infrastructures. 
Although conceptually similar to SAFARI, this work offers only an architectural
overview without detailing the enabling technologies required for
implementation. In contrast, our work focuses on technological specifics,
providing detailed guidance on implementing SAFARI and its practical
applications in experiments.

SCASS~\cite{DAMBROSIO2025104315}, although completely Open Source, proposes a DT oriented to the implementation of a specific Industrial Component System Use Case, which is composed by a mix of virtualized components and "Hardware-in-the-loop" physical components. This hybrid testbed is then utilized to test cyber attacks specific to the ICS context. 
Similar to SCASS, EPICTWIN~\cite{KANDASAMY2022108061} proposes an Open Source virtualization of a highly specific ICS Use Case and allows for live attack simulations on SCADA systems.
Differently from the last two cited works, SAFARI doesn't aim to reproduce a specific Use Case, but allows the user to configure any kind of virtualized network, will it be IT or OT/ICS oriented.

In terms of structure and implementation, one work close to SAFARI is
PANDORA~\cite{jiang2021pandora}, which is a safe testing environment that allows
users to conduct experiments on automated cyber-attack tools.
The differences between the two proposals lie in their focus. PANDORA is
designed mainly for testing automated cybersecurity tools, such as scanners or
IDS systems. In contrast, SAFARI focuses on creating test scenarios that are as
close as possible to real-world ones, prioritising open tools and highly modular networking 
virtualisation technologies, ensuring greater fidelity and enhancing flexibility in adapting to various testing needs.
We report the key points of the comparison in \cref{tab:summary_related}.

\newcommand{\thcel}[1]{\rotatebox{70}{#1}}
\begin{table}[t]
    \footnotesize
    \centering
    \begin{tabular}{c c c c c c c }
    \hline
    \thcel{\makecell{Compared solutions}} & \thcel{\makecell{Configurable Virtualized \\Infrastructure}} & \thcel{Sandboxing Capabilities} & \thcel{\makecell{General Purpose}} & \thcel{Hardware Agnostic} & \thcel{Air-gapped by design} & \thcel{Open Source Code}  \\[.5em]
    \hline
    \makecell{De Benedictis \\et al. \cite{10049521}} & \checker & \nocheck & \semicheck & \semicheck & \nocheck & \nocheck \\[.5em]
    \makecell{Masi et al.\cite{masi2023securing}} & \checker & \nocheck & \semicheck & \checker & \nocheck & \nocheck  \\[.5em]
     \makecell{PANDORA\cite{jiang2021pandora}} & \checker & \checker & \checker & \semicheck & \nocheck & \nocheck \\[.5em]
    \makecell{SCASS\cite{DAMBROSIO2025104315}} & \checker & \checker & \nocheck & \nocheck & \nocheck & \checker \\[.5em]
   \makecell{EPIC \cite{KANDASAMY2022108061}} & \checker & \checker & \nocheck & \semicheck & \nocheck & \checker \\[.5em]
   \hline
    \makecell{SAFARI} & \checker & \checker & \checker & \checker &
      \checker & \checker
    \end{tabular}   
\caption{Tabular comparison of SAFARI (last row) with related work (one work per
row) under the main characteristics of the considered proposals.}
\label{tab:summary_related}
\end{table}

%% file: conclusion.tex
\section{Conclusion}
\label{sec:conclusion}

Studying malware behaviour and testing detection and mitigation tools is
challenging, requiring strict safety protocols and reliable statistical data.
SAFARI addresses these needs by offering a democratised framework for ransomware
investigation, designed for scalability, air-gapped security, and automation,
catering to users including small research groups and educational institutions.
At its core, SAFARI is an open-source framework that automates ransomware (and
countermeasure) testing in realistic environments. Built with modern
technologies, it integrates Infrastructure-as-Code and OS-agnostic Task
Automation for reproducible and consistent experiment setups across diverse
hardware. We present a prototype that demonstrates SAFARI's feasibility and
effectiveness.

As an additional contribution, we conduct two case studies on profiling the
behaviour of five renowned ransomware strains (Ryuk, Vipasana, WannaCry,
LockBit, and Phobos) and evaluating the effectiveness of a ransomware mitigation
tool (Ranflood) against three of these strains. Thanks to SAFARI we described the distinct attack strategies of the strains. Briefly, Ryuk and Vipasana primarily target the ``AppData'' folder, leaving behind many ``.ignore'' files as part of their encryption approach, while WannaCry shows a broader attack pattern, affecting multiple user directories and targeting common file extensions. LockBit exhibits the most targeted strategy, focusing on user-important files such as those on the ``Desktop'', ``Documents'', and ``Pictures'' folders. Phobos seems the most aggressive ransomware, encrypting nearly all files indiscriminately. Studying the effectiveness of Ranflood, we found that the copy-based strategies are the most effective in preserving user-critical data.

Future work aims to expand SAFARI's capabilities to support a wider range of
malware types, such as exfiltration ransomware~\cite{MSLXWLHNH24}, by
incorporating tools to track malicious network activities. Another direction
involves enabling hybrid on-premises-cloud deployments, allowing users to scale
on-premises capacity with cloud resources. For the prototype, we plan to
integrate and automate additional analysis tools, such as those leveraging
dynamic ransomware behaviour analysis and automated malware behaviour pattern
recognition, to aid in interpreting results. To make SAFARI accessible to
non-technical users, we are exploring user-friendly interfaces to help define
high-level testing plans, which the framework would automatically translate into
corresponding IoC and OTA components.